# Harmonizing Material Quantity and Terahertz Wave Interference Shielding Efficiency with Metallic Borophene Nanosheets

Haojian Lin[†], Ximiao Wang[†], Zhaolong Cao, Hongjia Zhu, Jiahao Wu, Runze Zhan, Ningsheng Xu, Shaozhi Deng*, Huanjun Chen*, and Fei Liu*

State Key Laboratory of Optoelectronic Materials and Technologies, Guangdong Province Key Laboratory of Display Material and Technology, and School of Electronics and Information Technology, Sun Yat-sen University, Guangzhou 510275, China

[†]These authors contributed equally to the work.

*Corresponding authors' e-mail: liufei@mail.sysu.edu.cn; chenhj8@mail.sysu.edu.cn; stsdsz@mail.sysu.edu.cn.

**Materials with electromagnetic interference (EMI) shielding in the terahertz (THz) regime, while minimizing the quantity used, are highly demanded for future information communication, healthcare and mineral resource exploration applications. Currently, there is often a trade-off between the amount of material used and the absolute EMI shielding effectiveness ($EES_t$) for the EMI shielding materials. Here, we address this trade-off by harnessing the unique properties of two-dimensional (2D) $β_{12}$-borophene ($β_{12}$-Br) nanosheets. Leveraging $β_{12}$-Br's light weight and exceptional electron mobility characteristics, which represent among the highest reported values to date, we simultaneously achieve a THz EMI shield effectiveness (SE) of 70 dB and an $EES_t$ of $4.8 \times 10^5$ dB·cm$^2$/g (@0.87 THz) using a $β_{12}$-Br polymer composite. This surpasses the values of previously reported THz shielding materials with an $EES_t$ less than $3 \times 10^5$ dB·cm$^2$/g and a SE smaller than 60 dB, while only needs 0.1 wt.% of these materials to realize the same SE value. Furthermore, by capitalizing on the composite's superior mechanical properties, with 158% tensile strain at a Young's modulus of 33 MPa, we demonstrate the high-efficiency shielding performances of conformably coated surfaces based on $β_{12}$-Br nanosheets, suggesting their great potential in EMI shielding area.**

## Introduction

THz radiation, spanning from 0.1 to 10 THz in frequency, represents a frontier in technology with vast potential across multiple applications including information communications, radar imaging, noninvasive inspection, and biomedical sensing[1]. Its appeal lies in distinctive characteristics, such as low photon energy, robust penetration capabilities and the ability to discern characteristic molecular signatures within this spectral range. However, despite its promise, THz electronic and optoelectronic devices are not immune to electromagnetic interference (EMI) pollution, a challenge well-known in the radio-frequency realm[2-4]. Such interference can markedly degrade device performance and do harm to the surrounding environment, underscoring the need for effective EMI shielding materials.

An ideal EMI shielding material should effectively block electromagnetic waves (EMW) while simultaneously ensure that the isolated electromagnetic field does not emit into the surrounding environment. Meanwhile, such a material is also anticipated to be low-cost and easily producible, lightweight and possess good stretchability, enabling it to conform to arbitrary surfaces. This flexibility is particularly crucial for effective EMI shielding in wearable devices and portable equipment for consumer electronics.

Currently, research on THz EMW shielding materials follows two main approaches. On one hand, traditional materials originally designed for the microwave frequency range are still utilized, such as metal thin films[5-7] and carbon-based composite films, including conductive carbon inks[8,9], carbon fibers[9,10], carbon nanotubes[9,11], and graphene[9,12]. These materials are highly conductive and inhibit EMW through reflection, resulting in a significant portion of the waves being radiated into the surrounding environment, thus producing secondary pollution[5,13]. This inherent characteristic renders them less suitable for serving as EMI shielding materials in miniaturized devices with a demand of high integration. Alternatively, EMW can be shielded through absorption by the charged carriers within the EMI shielding materials. This has led to a recent shift in research focus towards exploring new-type materials with high electromagnetic absorption properties. These materials include polymer composite films[14,15], nanocomposite films[16,17], multilayer assemblies[5,15], porous foams/aerogel composites[17,18], and so on. Typically, these materials consist of segregated conductive fillers with a high density of free electrons with large mobility, facilitating the absorption of EMW through scattering at defects and interfaces presenting on each filler surface, followed by re-absorption by the free electrons. In particular two–dimensional (2D) conductive materials, such as MXene and graphene, have recently been employed as the fillers due to their outstanding electrically conductive characteristics. For example, a very recent study demonstrates that MXene assemblies can approach the intrinsic absorption limit in the 0.5–10 THz frequency range, with a small thickness of 10.2 nm[19]. A thin graphene/PMMA nanolaminate composite is displayed with a high conductivity to exhibit a shielding effectiveness (SE) reaching 60 dB for a small thickness of 33 μm[15]. Our recent study indicates that a transparent MXene film can have an SE of 21 dB over a wide frequency range of 0.1–10 THz[20].

While these exciting results provide a pathway for developing high-performance THz EMI shielding materials, there usually needs a trade-off between the mass amount of material used and the shielding effectiveness. This typically leads to an absolute EMI shielding effectiveness (EES$_t$) of less than $3 \times 10^4$ dB·cm$^2$/g and an SE smaller than 60 dB (Table S1, Supporting Information)[15]. To further enhance the shielding performance of these materials, an increased filler content or increment of the material's thickness is usually necessary. However, this inevitably increases the overall mass and thickness of EMI shielding material, which could compromise their mechanical stability and limit their developments in portable and miniaturized devices requiring lightweight, flexible, and conformable coatings. All these requirements highlight the necessity for a flexible shielding material with both high EES$_t$ and shielding SE values. Moreover, for practical applications, an environmentally favorable and scalable manufacturing process for the EMI shielding materials is urgently demanded, yet very challenging[2, 21].

Borophene, often termed the boron analogue of graphene, represents a unique class of 2D material characterized by its presence in multiple phases, stemming from the arrangement and deficiency of boron atoms[22, 23]. These phases encompass triangular, honeycomb, stripe and rectangular configurations, among others[24]. The diverse structural landscape of borophene, composed of the lightest solid element, underpins its ultralow density and distinguished flexibility, characterized by a high Young's modulus[22]. Furthermore, borophene is predicted to have outstanding electronic mobility alongside a large carrier density, suggesting it is one of the very few metallic or semi–metallic 2D materials with high conductivity[25, 26]. These properties render borophene an attractive candidate for embedding into polymer matrices to develop advanced flexible EMI shielding materials. However, the absence of a synthesis method for growing freestanding metallic borophene with high yield has hindered their developments on THz EMI shielding area, causing the low EES$_t$ smaller than 0.125 dB·cm$^2$/g and SE lower than 42 dB[27, 28], far from fully utilizing its advantages.

Here, we establish a scalable synthesis method for creating a wafer-scale composite of polydimethylsiloxane (PDMS) embedded with few-layer $\beta_{12}$-Br single crystalline nanosheets[29] as conductive fillers. The resulting composite film can reach a maximum size of up to 5 inches and has a thickness ranging from 0.2 to 4 mm. The single-crystalline $\beta_{12}$-Br fillers display superhigh-density free electrons (~$3.4\times10^{19}$ m$^{-2}$) and large conductivity. These free electrons are excited by THz illumination, dissipating through scattering at the boundaries of the $\beta_{12}$-Br nanosheets. This phenomenon significantly contributes to the exceptional performance of the composite film in THz EMI shielding. The film achieves a mean SE as high as over 70 dB and an ultrahigh

EES$_t$ of 4.8 × 10$^5$ dB·cm$^3$/g, the highest value to date as we know, across the 0.5 to 2 THz frequency range, with potential extension up to 10 THz. These results overwhelm those of previously reported EMI shielding materials designed for the THz spectral regime, which require at least 10$^4$ times more mass than the 2D borophene utilized in our current study to realize the comparable SE value. Additionally, the flexible composite film exhibits an ultrahigh tensile strain of over 158% at a Young's modulus of 33 MPa, enabling effective shielding of conformably coated surfaces. The results undoubtedly underscore the potential of few-layer metallic $β_{12}$-Br nanosheets as extremely efficient, low-density, and highly elastic materials for advanced THz shielding applications.

## Results and discussion

**Fabrication and characterization of $β_{12}$-Br/PDMS composites**

The $β_{12}$-Br comprises five atoms per unit cell, characterized by alternating rows of empty and filled hexagons along the $x$-direction. This arrangement usually produces stripes of vacancies along this axis. Along the $y$-direction, the structure consists of columns featuring a continuous line of atoms interspersed with incomplete hexagons, as shown in Fig. 1a. The unique crystalline structure of $β_{12}$-Br results in a high electron density (3.4 × 10$^9$/m$^2$), accompanied by a large electron mobility (2.84 × 10$^6$ cm$^2$/(V·s)) and a high Young's modulus (382 GPa) [24,25]. Furthermore, it is the most stable among the various allotropes of borophene in its freestanding state[23, 29-32]. In this study, the borophene nanosheets were synthesized by our developed low-temperature liquid phase exfoliation (LTLE) technique (see Methods for details)[33], as depicted in Fig. 1b. The crystalline structure and stoichiometric ratio of the products are respectively confirmed by X-ray diffraction (XRD) and confocal Raman spectroscopy, revealing the nanosheets are the $β_{12}$ phase with high crystallinity (Supplementary Figs. 1 and 2). The $β_{12}$-Br nanosheets are a few atomic layers in thickness and well-dispersed into an aqueous solution, with an average thickness about 4 nm (Fig. 1c, Supplementary Fig. 3) and a six-fold symmetry (Fig. 1d, Supplementary Fig. 4). Additionally, over 94 at.% of the boron atoms remain unoxidized (Supplementary Figs. 5 and 6). The average sheet resistance of the $β_{12}$-Br nanosheets is determined to be 5.6 × 10$^4$ Ω/sq (Supplementary Section 3 and Supplementary Fig. 7), suggesting their metallic nature. This resistance value is approximately five orders of magnitude smaller than that previously reported for the mixture phase of $\beta_{12}$-Br and $\chi_3$-Br[34, 35], underscoring the superior crystallinity and purity of the $\beta_{12}$-Br synthesized using our method. This high conductivity is further supported by electrical conductivity measurement from a typical $\beta_{12}$-Br nanosheet with a thickness of 4 nm, yielding a value of $2.6 \times 10^{-5}$ Ω·m (Supplementary Section 3 and Supplementary Fig. 8), unveiling $\beta_{12}$-Br nanosheets should be suitable for THz wave EMI shielding.

It should be noted that fabricating robust, stretchable, and continuous thin film based on pure borophene nanosheets alone has significant challenges, especially when attempting to scale up to larger area[27]. Instead, we employ the $\beta_{12}$-Br nanosheets as the fillers embedded into PDMS film to form a flexible composite. Specifically, the composite film was produced using a simple sol-gel method (Fig. 1b), where the $\beta_{12}$-Br nanosheets were intricately linked together *via* robust intermolecular bonding interactions involving hydrogen atoms from PDMS molecules and boron atoms located on the surface of the $\beta_{12}$-Br (Fig. 1a). This approach has been previously employed, where $\alpha$-Br flakes were used as fillers[27]. However, due to the inefficient EMI shielding of $\alpha$-Br reported in those studies, elevated concentration of borophene, up to 100 wt.%, was necessary. This undoubtedly causes the difficulties in achieving uniform and continuous thin films owe to potential cracking of the PDMS matrix. For the $\beta_{12}$-Br, as discussed below, leveraging the ultrahigh $EES_t$ of the $\beta_{12}$-Br allows for exceptionally low borophene content of less than 0.5 wt.% in the matrix and minimizing the breakage probability of PDMS intermolecular bond. This obvious decrease of the $\beta_{12}$-Br content facilitates their homogeneous dispersion within the PDMS matrix (Fig. 1f, inset), enabling the formation of flexible large-area thin films up to 5 inches (Fig. 1e, f).

The THz wave shielding performance of the $\beta_{12}$-Br/PDMS composite film was evaluated using a THz time–domain spectroscope (THz–TDS) system. With a thickness of 2 mm and containing 0.13 wt.% of $\beta_{12}$-Br, the EMI SE of the composite film can reach as high as 70 dB (Fig. 1g). Furthermore, the EMI $EES_t$, which offers a more comprehensive assessment of the material's shielding performance under idealized conditions by considering the mass of the shielding materials and the spot size of incident EMW (see Methods), is obtained to be in the range of $2.5 \times 10^5 - 4.8 \times 10^5$ dB·cm$^2$/g across the frequency range of 0.8 THz to 2 THz. Notably, several films, comprising of graphene or MXene composites among others, demonstrate the EMI $SSE_t$ maxima capped at $3 \times 10^5$ dB·cm$^2$/g. However, these composite materials, even at concentrations of at least 5 wt.%, fell short of achieving the benchmark of 80 dB in terms of EMI SE (Fig. 1h, Table S1, Supporting Information). In stark contrast, the $\beta_{12}$-Br/PDMS film, comprising a mere 0.13 wt.% weight fraction and measuring 2 mm in thickness, showcased an impressive average EMI $SSE_t$ of $4.8 \times 10^5$ dB·cm$^2$/g. Furthermore, with an increase in the weight fraction to 0.5 wt.% and the thickness to 4 mm, the film attained an average EMI SE as high as 85 dB. Such comparison evidently underscores that despite of the evidently improved performance of the $\beta_{12}$-Br/PDMS film, the mass of the shielding materials is four orders of magnitude smaller than that of the previous materials.

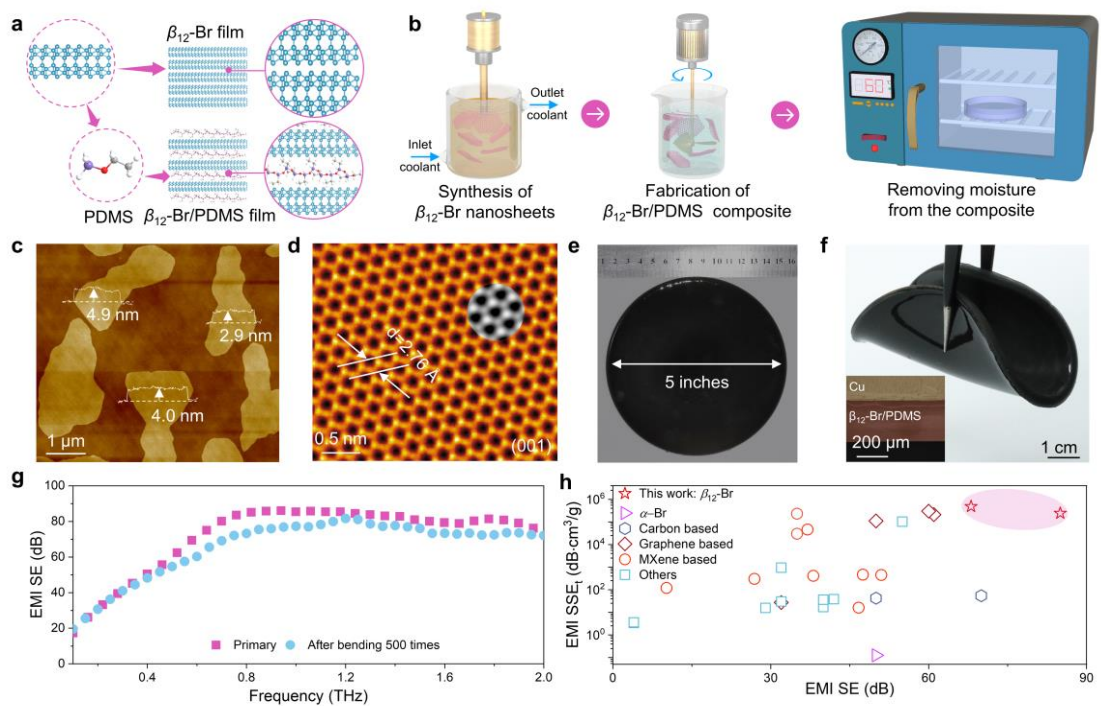

**Fig. 1 | Characterization and THz shielding measurements of $\beta_{12}$-Br/PDMS composite film. a,** A schematic diagram illustrating the preparation process of $\beta_{12}$-Br nanosheets and $\beta_{12}$-Br/PDMS composite film. **b,** The production route for $\beta_{12}$-Br/PDMS composite film. **c,** An AFM image displaying the topography of few-layer $\beta_{12}$-Br nanosheets synthesized by LTLE way, where the monolayer thickness is indicated by the inserted height profile. **d,** A typical HRTEM image of a $\beta_{12}$-Br nanosheet, with the inset presenting the theoretical surface configuration of borophene (001) face using density functional theory (DFT). **e,** Representative photograph showcasing the as-synthesized centimeter-scale $\beta_{12}$-Br/PDMS composite film. **f,** A photograph demonstrating the excellent flexibility of $\beta_{12}$-Br/PDMS composite film, with an inset showing the corresponding cross-section image. **g,** The frequency-

dependence analysis of EMI SE curves of the 4-mm-thickness composite film with a 0.5 wt.% $β_{12}$-Br nanosheets before and after undergoing 500 bending cycles. **h,** Comparison between EMI SE, and EMI $SSE_t$ values obtained from $β_{12}$-Br/PDMS film and other excellent shielding materials.

**THz shielding performance and mechanism of $β_{12}$-Br/PDMS composite film**

A series of $β_{12}$-Br/PDMS composite films were prepared to investigate the influence of the weight ratio (wt.%) of $β_{12}$-Br nanosheets and the film thickness on the THz EMI shielding performance (Supplementary Fig. 9). When 1.09 wt.% bulk boron powders were added into the 2-mm-thickness PDMS film, the maximum EMI SE value only increases slightly to 29 dB (Fig. 2a, blue). Instead, if the filler is changed to the $β_{12}$-Br nanosheets, the THz shielding performances of the composite film were significantly improved (Fig. 2a). For instance, in a composite film with 0.13 wt.%, the SE already reaches as high as 68 dB at 0.87 THz. Furthermore, the EMW shielding behavior extends across a wide frequency range from 0.1 to 8 THz (Supplementary Fig. 10), highlighting that the $β_{12}$-Br/PDMS composite film serves as an ultra-broadband EMW shielding material.

Through simultaneous measurement of the reflectance and transmittance spectra of the composite film, its EMI absorption effectiveness ($SE_A$) and reflection effectiveness ($SE_R$) can be obtained, respectively. Especially, for the illumination frequency larger than 0.2 THz, the average $SE_A$ reaches up to 60 dB while the $SE_R$ is only ~5 dB (Supplementary Fig. 11). The results clearly show that the incident THz electromagnetic wave almost experiences none of the reflection and transmission loss. Instead, the power of THz EMW is effectively dissipated by the composite film, specifically by the $β_{12}$-Br nanosheets. This is further elucidated by monitoring the SE spectra measured from composite films with varying concentrations of $β_{12}$-Br nanosheet but identical thickness (2 mm). By progressively increasing the borophene content from 0.13 to 2.1 wt.%, the EMI SE initially experiences rapid augmentation, eventually reaching a plateau for mass ratios over 0.5 wt.% (Fig. 2b). Notably, the highest SE achieved can arrive at 76 dB at 0.87 THz for a mass ratio of 2.1 wt.%, unveiling that the EMW shielding of the composite film is primarily attributed to the addition of $β_{12}$-Br nanosheets.

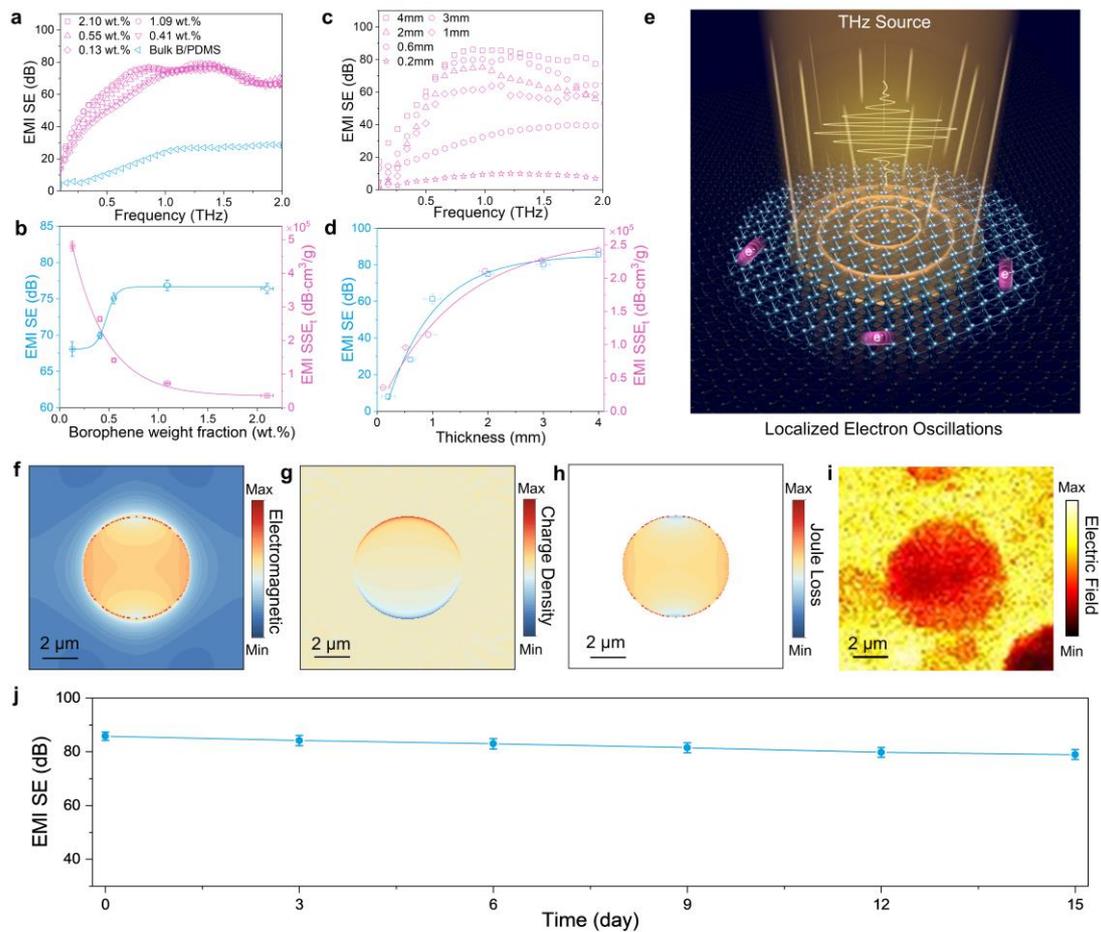

**Fig. 2 | EMI shielding properties of $\beta_{12}$-Br/PDMS composite film. a**, The EMI SE curves of the composite film with various weight fractions of $\beta_{12}$-Br nanosheets. **b**, The curves of EMI SE/EMI SSE$_t$ to the weight fraction of borophene nanosheets at 0.87 THz. **c**, The frequency-dependent EMI SE curves of the composite film with different thickness. **d**, The relationship between EMI SE/EMI SSE$_t$ and film thickness at 0.87 THz. **e**, Schematic showing the excitation of electron oscillations in $\beta_{12}$-Br nanosheets under THz radiation. **f–h**, Characterization of the simulated charge density, joule loss and energy density of individual $\beta_{12}$-Br nanosheet in the frequency range of 0.1 – 2 THz. **i**, The THz near-field image of a single $\beta_{12}$-Br nanosheet on a SiO$_2$ wafer. **j**, The EMI SE stability of the $\beta_{12}$-Br/PDMS composite film after 15 days' air storage.

The aforementioned results clearly show that even a small (2.1 wt.%) addition of $\beta_{12}$-Br nanosheet into the polymer matrix can yield a superhigh EMI SE of 76 dB. This obviously surpasses previous findings, where an SE of only 42 dB was achieved with 100 wt.% of $\alpha$-Br filled into the same matrix[27]. To further highlight the outstanding EMW shielding performance of the $\beta_{12}$-Br/PDMS composite film, the dependence of EMI SSE$_t$ on the mass ratio of borophene nanosheets is characterized, revealing a nonlinear increase with the reduction of the weight ratio (Fig. 2b). Most of all, for a weight ratio of 0.13 wt.%, the EMI SSE$_t$ can reach up to $4.8 \times 10^5$ dB·cm$^3$/g, superior to all reported composite films with conductive fillers such as graphene and MXene (Fig. 1h). Simultaneously, the EMI SE value of $β_{12}$-Br/PDMS composite film can still maintain as high as 68 dB, which achieves the best value reported up to date.

The THz wave shielding performance of the composite film is strongly dependent on the film thickness, as manifested from the EMI SE spectra against the film thickness (Fig. 2c). If the weight ratio of borophene nanosheets was kept at 0.5 wt.%, both of the EMI SE and SSE$_t$ values monotonically increase with the film thickness (Fig. 2d). Particularly, for a film thickness of 4 mm, the maximum EMI SE and SSE$_t$ can reach as high as 85 dB and $2.5 \times 10^5$ dB·cm$^2$·g$^{-1}$ at 0.87 THz, respectively. These results evidently demonstrate the exceptional EMW shielding performance of the $β_{12}$-Br/PDMS composite film.

Afterwards, the mechanism governing the strong THz EMW absorption of the $β_{12}$-Br/PDMS composite film are further explored. The preceding discussion unambiguously suggests that the incident EMW are efficiently absorbed and dissipated by various $β_{12}$-Br nanosheets within the composite film. Each single crystalline $β_{12}$-Br nanosheet supports a high concentration of free electrons with large mobilities. Upon excitation by the THz wave, these free electrons become excited and accelerated by the electric field of the EMW, leading to collective oscillations. However, due to the small planar size of the borophene nanosheets (approximately 3 μm, Fig. 1c), the electrons encounter boundaries and suffer from multiple reflections, ultimately losing their kinetic energies into the lattice of the $β_{12}$-Br nanosheets, as seen in Fig. 2e. In this manner, nearly all of the absorbed EMW are dissipated as heat, resulting in the ultrahigh EMI absorption effectiveness observed.

This mechanism can be further supported by simulating the dynamics of electrons and the corresponding absorption of EMW energy within a $β_{12}$-Br nanosheet (Methods and Supplementary Section 5). The behavior of free electrons in $β_{12}$-Br nanosheet in response to the EMW is governed by the alternating-current conductivity, which can be characterized using the Drude model (Supplementary Figs. 12 and 13). For a disk-shaped sheet with a monolayer thickness, multiple electromagnetic resonances, characterized by strong absorption peaks, can be excited upon irradiation by the THz wave (Supplementary Fig. 14). As shown in Fig. 2f, these resonances induce strong EMW localizations in $\beta_{12}$-Br nanosheets, where the associated free electrons become localized and subsequently reflected by the disk boundary (Fig. 2g). Accordingly, lots of Joule loss will occur within the borophene nanosheets (Fig. 2h), which can be further confirmed by THz near-field optical measurements (Method) conducted on an individual $\beta_{12}$-Br nanosheet (Fig. 2i). As a result, the Joule loss leads to a very strong absorption efficiency of up to 50% at the resonance, even for a nanosheet with a monolayer thickness (Supplementary Fig. 14).

It is noteworthy that, according to the simulation results, the resonance frequency of the $\beta_{12}$-Br nanosheet is strongly dependent on its size and shape, covering a broad spectral range, for example, from 0.2 to 2 THz (Supplementary Fig. 14). However, the maximum absorption efficiency nearly remains unvaried at 50%, irrespective of the geometrical parameters of the borophene nanosheets (Supplementary Fig. 14). Consequently, the broadband yet relatively steady THz EMI absorption shielding performance of the $\beta_{12}$-Br/PDMS composite film observed in our study should originate from the containing $\beta_{12}$-Br nanosheet with varied thickness and shapes in the composite film.

Interestingly, even after being exposed to air over 15 days, the EMI shielding performances of the $\beta_{12}$-Br/PDMS composite film showed no obvious degradation and retained 92.94% of the original performance (Fig. 2j), revealing the high stability of the $\beta_{12}$-Br/PDMS film under ambient conditions.

**Mechanical properties of the $\beta_{12}$-Br/PDMS composite film**

Good mechanical property is another critical requirement for THz EMW shielding materials in portable and wearable electronic devices[5, 36]. By taking advantage of the exceptional Young's modulus of $\beta_{12}$-Br (theoretical value of 163 to 382 GPa)[37], the composite film can be easily bent up to a high angle of 180° without any breakage or cracks observed through 500 bending experiments, revealing its excellent flexibility. Further uniaxial tensile experiments were conducted to quantitatively assess the elastic behavior of the composite film with various weight fractions of borophene nanosheets. It is noted that the 0.13 wt.% $\beta_{12}$-Br/PDMS composite film exhibits a tensile strain, $\sigma_s$, of 50% at a tensile stress, $\varepsilon_s$, of 13 MPa, while that with 2.10 wt.% achieves a $\sigma_s$ of 158% at $\varepsilon_s$ = 33 MPa (Fig. 3a). The corresponding Young's moduli, $E_c = \sigma_s/\varepsilon_s \times 100$,[38] were approximately 26 MPa and 21 MPa, respectively. One can obviously see that the mechanical properties of the composite film are gradually enhanced with increasing the weight fraction of borophene, overwhelming those of many other high-modulus 2D materials (Table S2, Supporting Information). Also, this observed trend corroborates that the ultrahigh theoretical Young's modulus of borophene nanosheets should be responsible for the excellent mechanical properties of the composite film.

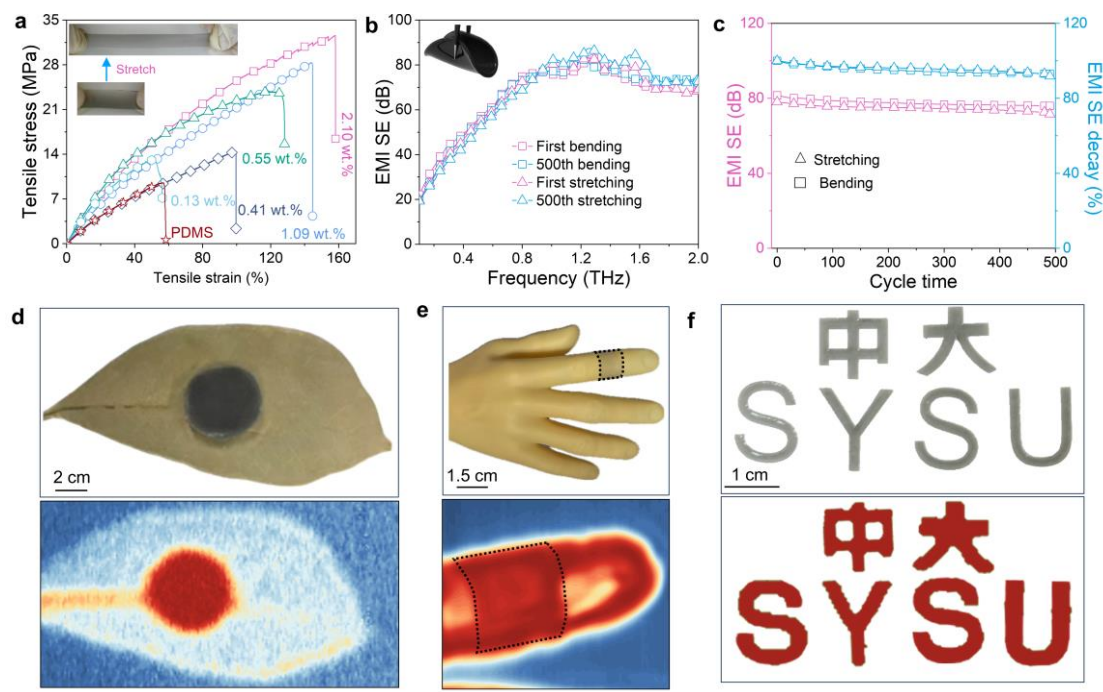

**Fig. 3 | Mechanical properties of $β_{12}$-Br/PDMS composite films and their THz shielding applications. a**, Representative stress-strain curves of the 2-mm-thickness composite film with different borophene contents in uniaxial tensile test. The inset at the top left corner shows a photograph of the composite film undergoing tensile testing. **b**, EMI SE values of the composite film after 500 times' bending or stretching measurements. The photograph of the composite film undergoing bending is displayed in the inset. **c**, The EMI SE value and EMI SE decay ratio of the composite film at 0.87 THz during 500 bending or stretching cycles, respectively. **d**, **e**, Digital images of the leaf coating and the human finger coated by the planar and curved composite film, respectively, in which the THz wave is effectively shielded and the inner information can't be differentiated. **f**, THz imaging of the Sun Yat-sen University masked by shaped $β_{12}$-Br/PDMS composite film.

THz shielding measurements were then performed on the 2-mm-thickness composite film with 0.5 wt.% $β_{12}$-Br nanosheets. After 500 bending cycles, the mean EMI SE value of the film slightly decreased from 81 dB to 75 dB, with an average fading rate as low as 0.015% per bending cycle (Fig. 3b). Similarly, the mean THz EMI SE value of the composite film mildly reduced from 78 dB to 71 dB after 500 stretching cycles, with an average fading rate of less than 0.018% per stretching cycle (Fig. 3c). Despite of these tiny attenuation, the sample can recover over 80% of its initial EMI SE efficiency for THz waves after both bending and stretching tests. These findings validate the great potential of $\beta_{12}$-Br/PDMS composite film to combine high flexibility with superior mechanical strength, making them promising candidates for advanced flexible electronic applications.

**Application demonstration of the $\beta_{12}$-Br/PDMS composite film**

The discussion above clearly exhibits the excellent THz shielding performance and flexibility of the $\beta_{12}$-Br/PDMS composite film, which ensure them highly suitable for applications in EMW shielding of objects with irregular surfaces. To this end, the $\beta_{12}$-Br/PDMS composite film was employed to conformably coated onto practical objects to show its excellent EMI SE upon THz EMW illumination. A 1-mm-thickness $\beta_{12}$-Br/PDMS composite film was placed on the surface of a dry leaf for THz shielding imaging (Fig. 3d). As observed in Fig. 3e, the THz signal is almost completely absorbed where the composite film covers the leaf, causing the emergence of the shadow for imaging leaf. Additionally, the same composite film was wrapped around a human finger, effectively blocking off the THz signal at the wrapped region. Moreover, the composite film can also be fabricated into different patterns to match the outlines of target objects for demonstrating different characters, as presented in Fig. 3e. It is obviously seen that the contour profile of "Sun Yat-sen University" is very clear and sharp, further showcasing the versatility of the composite film in being processed into EMW shielding materials with arbitrary shapes to meet diverse application requirements.

# Discussion

Borophene, an important member of 2D material family, has attracted much attention since the birth due to its unique electrical and mechanical properties, as well as low mass density. While their applications in energy conversion and storage have been widely studied, their photonic and optoelectronic applications remain unexplored. The main obstacle lies in that it is still a challenging issue to produce freestanding borophene with high yield and high purity out of a variety of different phases. In this study, we demonstrate the successful applications of borophene in THz EMW shielding by developing a facile approach for creating composite film consisted of PDMS filled with single-phase and high-crystallinity few-layer $\beta_{12}$-Br nanosheets of high conductivity. Such composite film, which can be scaled up to 5 inches and with tailorable thickness, exhibits an ultrahigh SE and EESt over 70 dB and $4.8 \times 10^5$ dB·cm$^3$/g, respectively, with only 0.13 wt.% filling ratio of the $\beta_{12}$-Br nanosheets. This successfully circumvents the trade-off between the amount of filling material used and the shielding effectiveness, which usually present in the previous reported EMI shielding materials designed for the THz spectral regime.

Moreover, the remarkable mechanical property of single crystalline $\beta_{12}$-Br nanosheets guarantees the composite film not only handily conforms to different object surfaces with various three-dimensional curvatures, but also produces efficient shielding on these objects. One possible application for our flexible composite film is to prevent the EMW pollution in electrical and optoelectrical circuits from the surrounding environment or adjacent neighborhoods in future 6G communication networks or large-scale integrated circuits. Besides, with its low-weight, high-efficiency, good-flexibility, and high-stability, and broadband shielding performances, the composite film based on $\beta_{12}$-Br nanosheets could be successfully utilized in a wearable device for daily consumer electronics in the future.

## Methods

**Materials.** Boron powders (99.8%, 325 mesh) were purchased from ZhongNuo (China) Co. Ltd. Polydimethylsiloxane (PDMS, SYLGARD 184) and N-methyl pyrrolidone (NMP, 99.8%) was bought from Dow–Corning (USA) Co. Ltd. and Innochem (China) Co. Ltd, respectively.

**Sample Fabrication.** $\beta_{12}$-Br nanosheets can be synthesized using our previously developed LTEP method[33]. And PDMS gel was prepared by mixing the base component with the hardener in a weight ratio of 10:1.

The procedures to fabricate $\beta_{12}$-Br/PDMS composite film are as follows. Firstly, the as-grown $\beta_{12}$-Br nanosheet powders were dissolved into 1 mL deionized water via 30 min of bath ultrasonic dispersion, and subsequently added into the PDMS solvent to form the uniform sol by a 1 hour of continuous stirring. Secondly, the sol was coated on the surface of laboratory dish and sat for 2 days in air to realize the solidification of the gel. Finally, they were handed into the vacuum chamber and treated at 60 °C for about 12 hours to remove the residual moisture. Through the above procedures, the fabrication of large-area $\beta_{12}$-Br/PDMS composite film was accomplished, as shown in Fig. 1b.

**Characterization.** The thickness of $\beta_{12}$-Br nanosheets was measured by an atom force microscope (AFM, Bruker Dimension Icon). The morphology and crystalline structure of the nanosheets were investigated by a scanning electron microscope (SEM, Zeis Supra 60) and transmission electron microscope (TEM, FEI Titan 80-300). The scanning transmission electron microscope (STEM), energy disperse X-ray (EDX) mapping and electron energy loss spectroscopy (EELS) techniques were carried out in a JEM ARM200F thermal-field emission microscope with a Cs corrector probe working at 300 kV. For the high-angle annular dark field (HAADF) measurement, a convergence angle of about 21 mrad and collection angle range of 65-172 mrad were adopted for the incoherent atomic number imaging. The chemical compositions were analyzed by XRD patterns recorded on a D-MAX 2200 VPC system. The Raman spectrum of the $\beta_{12}$-Br nanosheets was obtained by inVia Reflex (532 nm laser) made by Renishaw. And the current-voltage characteristics of the borophene nanosheet were tested in an ultra-high vacuum (UHV) probe made by Wavetest.

**Computational model of $\beta_{12}$-Br nanosheet.** The theoretically model of few-layer $\beta_{12}$-Br nanosheet was obtained by the density functional theory (DFT), and more details can be found in our previous report[33].

**THz Shielding Measurements.** The THz shielding effectiveness of the samples was studied using a THz-TDS (Toptica) system at room temperature under $N_2$. The samples were attached onto a hollow iron plate for test, and THz wave focused on the sample with a spot radius of 2.5 mm. The EMI SE of the material can be described by decibels (dB) and derived using the following equation:[5]

$$\text{EMI SE (dB)} = -20\log_{10}\left(\frac{E_{in}}{E_{out}}\right) \quad (1)$$

, where $E_{in}$ and $E_{out}$ denote the incident field strength and transmitted field intensity of THz waves, respectively. The EMI SSE$_t$ of the material can be calculated based on the following equation:[15]

$$\text{EMI SSE}_t \ (\text{dB}\cdot\text{cm}^2\cdot\text{g}^{-1}) = \frac{\text{EMI SE} \cdot S^2}{m} \quad (2)$$

, where $S$ and $m$ respectively represent the area of the focal spot of THz wave and the mass of $\beta_{12}$-Br nanosheets.

**THz Optical Nanoimaging.** Optical nanoimaging was conducted on the samples using a scattering-type THz optical microscope (THz-NeaSNOM, Neaspec GmbH). To image the $\beta_{12}$-Br nanosheet in real space, a THz laser with tunable frequency from 0.1 to 3 THz was focused onto both the sample and a metal-coated AFM tip (25PtIr200B-H, Rocky Mountain Nanotechnology) with a radius below 20 nm. The back-scattered light from the tip was demodulated and detected at a harmonic higher than that of the tip.

**Numerical Simulation.** Because $\beta_{12}$-Br nanosheet belongs to a metallic 2D material, its conductivity ($\sigma_{jj}$) can be obtained by the Drude model and written as:[39, 40]

$$\sigma_{jj} = \frac{iD_j}{\pi\left(\omega+\frac{i}{\tau}\right)}, \ D_j = \frac{\pi e^2 n}{m_j} \quad (3)$$

, where $j$ represents the $x$ or $y$ direction of the optical axis of $\beta_{12}$-Br nanosheet in our paper. In Eq. (3), $e, n, \omega, \tau, D_j$ and $m_j$ stand for electron charge, density of electrons, frequency of excitation, carrier lifetime, Drude weight along x or y direction, and effective electron mass in $x$ and $y$ directions, respectively. Therefore, the real ($\varepsilon_{r,jj}$) and imaginary ($\varepsilon_{i,jj}$) parts of the complex permittivity along each direction can be derived based on the following equation:

$$\varepsilon_{r,jj} = \varepsilon_r - \frac{e^2 n}{m_j \varepsilon_0 h \left(\omega^2 + \frac{1}{\tau^2}\right)}, \ \varepsilon_{i,jj} = \frac{e^2 n / \tau}{m_j \varepsilon_0 h \omega \left(\omega^2 + \frac{1}{\tau^2}\right)} \quad (4)$$

, where $\varepsilon_r = 11$ is the relative permittivity, $\varepsilon_0 = 8.854 \times 10^{-12}$ F/m is the vacuum permittivity, and $h$ represents the thickness of $\beta_{12}$-Br nanosheet. The effective electron mass (m$_x$/m$_y$) of $\beta_{12}$-borophene nanosheet along $x$ or $y$ directions are respectively 3.5 m$_0$ and 3.7 m$_0$, where $m_0 = 9.11 \times 10^{-31}$ kg is the mass of electron[25].

Using Matlab R2021b software, the permittivity spectra in the frequency range of 0.1–2 THz of $\beta_{12}$-Br nanosheets can be calculated according to Eqs. (3) and (4), where the carrier density ranges from $1.0 \times 10^{19}$ to $5.0 \times 10^{19}$ m$^{-2}$ and the mean free time of electrons varies from $3 \times 10^{-14}$ to $12 \times 10^{-14}$ s. And then, the permittivity spectra were imported into CST software to construct a $β_{12}$-Br nanosheet. Finally, we carried out the finite-difference time-domain simulations (FDTD, Lumerical Inc.) to simulate the THz absorbance of $β_{12}$-Br nanosheets.

## Data availability

The authors declare that the main data supporting the findings of this study are available within the paper. Extra data are available from the corresponding authors upon reasonable request. Source data are provided with this paper.

## Acknowledgements

The authors are very thankful for the support of the National Science Foundation of China (Grant No. 51872337), National Science Foundation of Guangdong Province (Grant No. 2021A1515012592) and the Science and Technology Department of Guangdong Province (Grant No. 2020B1212060030).

## Author contributions

S. D., H. C. and F. L. proposed and supervised the projects. H.L. synthesized the materials, characterized their surface morphology and chemical compositions, and carried out the THz shielding performances measurements. Z.C. and H.Z. simulated the absorption spectra of borophene in THz band by FTDT software. X.W. was responsible for the terahertz imaging of the $β_{12}$-Br/PDMS composite film. J.W. created a flowchart illustrating the experimental process. All the authors involved in the analysis and discussion of the experimental results. And all authors approve to submit the final version of the manuscript.

## Online content

Any methods, additional references, Nature Research reporting summaries, source data, extended data, supplementary information, acknowledgements, details of author contributions and competing interests; and statements of data and code availability are available at https://doi.org/xxxx.

## Competing interests

The authors declare no competing interests.

## Additional information

**Supplementary information** The online version contains supplementary material available at https://doi.org/xxxx.

**Correspondence and requests for materials** should be addressed to Fei Liu, Huanjun Chen or Shaozhi Deng.

**Peer review information Nature Materials** thanks xxx reviewer(s) for their contribution to the peer review of this work.

**Reprints and permissions information** is available at www.nature.com/reprints.